# A magnetic structure of ruthenium dioxide ($RuO_2$) and altermagnetism

S. W. Lovesey [1,2,3], D. D. Khalyavin,[1] and G. van der Laan[2]

[1]*ISIS Facility, STFC, Didcot, Oxfordshire OX11 0QX, United Kingdom*
[2]*Diamond Light Source, Harwell Science and Innovation Campus, Didcot, Oxfordshire OX11 0DE, United Kingdom*
[3]*Department of Physics, Oxford University, Oxford OX1 3PU, UK*

**Abstract** The magnetic structure of $RuO_2$ and the Ru atomic configuration are unknown. A magnetic structure is inferred by confronting measured and calculated Bragg diffraction patterns and adjusting the latter to achieve satisfactory agreement. An accepted pattern, a magnetic symmetry, includes symmetry of sites occupied by the magnetic ions. As a realistic starting point, we provide diffraction patterns for a magnetic symmetry of $RuO_2$, a descendent of the tetragonal parent structure, which accommodates a departure of Ru axial dipoles from the crystal c axis. A chiral signal and piezomagnetic effect are permitted, and a linear magnetoelectric effect forbidden. Features of the neutron diffraction pattern test the non-relativistic requirement of altermagnetism, and we scrutinize published room-temperature data. Specifically, one Bragg point is consistent with Ru orbital angular momentum and magnetic quadrupole both zero, and the latter result is not expected from non-relativistic altermagnetism. Azimuthal angle scans in resonant x-ray diffraction are sensitive to the Ru site symmetry and the atomic configuration. Acid tests of the studied magnetic symmetry include a chiral signature and null intensity for unrotated photon polarization.

*Introduction*. Ruthenium dioxide is a prominent face of altermagnetism, yet there is no consensus on its magnetic structure or the Ru electronic configuration [1, 2]. We frame pertinent questions using a realistic magnetic symmetry, and pose some answers available from diffraction experiments. Altermagnetism is a non-relativistic theory dedicated to collinear magnetic structures with perfect translation invariance. At face value, the theory is relevant to materials with a negligible spin-orbit coupling in the electronic configuration, and a magnetic structure with a null propagation vector. Spin degrees of freedom are completely decoupled from the lattice in the non-relativistic space groups of altermagnetism, which include transformations of spins and lattice that are not symmetry elements if one takes account spin orbit coupling. In consequence, symmetry tools embedded in altermagnetism must be applied consistently to avoid pitfalls from a mix and match theory.

The magnetic axial quadrupole in neutron diffraction is zero if the magnetic atomic state is a single J-state ($\mathbf{J} = \mathbf{S} + \mathbf{L}$ is the total angular momentum), as in the extreme $j_{eff}$ model of iridates with strong spin-orbit coupling [3]. Thus, the observation of a significant magnetic quadrupole in $RuO_2$ would weigh in favour its description by altermagnetism. Similarly, resonant x-ray diffraction is an excellent probe of local charge-like and magnetic angular anisotropy. According to Neumann's Principle, such anisotropies delineate the local symmetry of resonant ions [4, 5]. Thermal transport coefficients calculated for a model of $RuO_2$ show a striking dependence on the Néel vector derived from neighbouring Ru ions in a rutile structure. Specifically, this concerns the temperature dependent anomalous Nernst (thermoelectric), thermal Hall and Hall conductivities [6]. However, Zhou *et al*. [6] find null values for all the transport coefficients for a Néel vector parallel to the crystal c axis, which is demanded by rutile magnetic symmetry ($P4_2'/mnm'$). By adopting a magnetic symmetry for $RuO_2$ descended from the tetragonal (rutile) crystal structure, the Néel vector is not constrained to the c axis. We give a coherent account of neutron and x-ray diffraction patterns derived from the reduced symmetry to be tested in future experiments.

Magnetic symmetry $P2_1/c$ chosen for $RuO_2$ does not break translation symmetry and the propagation vector $\mathbf{k} = (0, 0, 0)$. It accommodates an antiferromagnetic motif of axial dipole moments in the tetragonal (bc) plane together with a ferromagnetic component along the a axis, as depicted in Fig. 1 (for structure information see Supplemental Material [7]). A linear magnetoelectric effect is forbidden, because anti-inversion is absent in the $P2_1/c$ crystal class, and a piezomagnetic effect is allowed. Notably, magnetic symmetry $P2_1/c$ responds to helicity in a beam of photons.

In the following, magnetic properties are referred to orthogonal vectors labelled ($\xi$, $\eta$, $\zeta$) derived from the monoclinic unit cell depicted in Fig. 1. The unique axis $\eta$ is parallel to the tetragonal a axis. Conventionally, the development of magnetic order leads to a lowering of the symmetry in the sample and the magnetic ordering pattern can be inferred by confronting experimental patterns derived by neutron diffraction with a symmetry analysis in which selected elements of crystal symmetry are assumed to have disappeared. Our results for neutron and x-ray diffraction patterns are for Bragg spots forbidden by the core structure, often labelled basis forbidden reflections. By construction, Bragg spots contain only magnetic and aspherical charge-like electronic contributions. Reflection vectors for tetragonal and monoclinic structures are labelled $(H_o, K_o, L_o)_t$ and $(h, k, l)_m$, respectively.

*Magnetic quadrupole*. A dependence of the magnetic neutron scattering amplitude on both the magnitude and direction of the reflection vector $\boldsymbol{\kappa}$ is a most valuable property of the technique. It enables the measurement of the magnetization density, or its spatial Fourier transform more correctly, and the identification of electron back-transfer (covalency) [9, 10]. Ruthenium dioxide can be described as a strongly covalent intermediate coupling system, and the Ru atomic configuration is defined by a mixture of J-states.

The axial dipole $\langle \mathbf{t}^1 \rangle$ contains standard radial integrals $\langle j_0(\kappa) \rangle$ and $\langle j_2(\kappa) \rangle$ depicted in Fig. 2 for the atomic configuration $4d^4$, with $\langle j_0(0) \rangle = 1$ and $\langle j_2(0) \rangle = 0$ [11, 12]. An approximation to $\langle \mathbf{t}^1 \rangle$,

$$\langle \mathbf{t}^1 \rangle \approx (2/3) \, [\langle j_0(\kappa) \rangle \, \langle \mathbf{S} \rangle + (1/2) \, (\langle j_0(\kappa) \rangle + \langle j_2(\kappa) \rangle) \, \langle \mathbf{L} \rangle], \qquad (1)$$

is often used [10, 13]. The numerical coefficient of orbital angular momentum in the ground state $\langle \mathbf{L} \rangle$ is approximate, while $\langle \mathbf{t}^1 \rangle = (1/3) \, \langle 2\mathbf{S} + \mathbf{L} \rangle$ for $\kappa \rightarrow 0$ is an exact result. Eq. (1) implies $\langle \mathbf{t}^1 \rangle = 0$ for $d^4$, because the atomic configuration is a singlet $J = 0$. One finds a magnetic moment

$\langle \mu_\zeta \rangle = \langle (2\mathbf{S} + \mathbf{L})_\zeta \rangle = -\langle \mathbf{L}_\zeta \rangle = \chi(1)_0 \sqrt{2}$, $\langle t^1_\zeta \rangle = (1/3) \langle \mu_\zeta \rangle [\langle j_0(\kappa) \rangle - (3/4) \langle j_2(\kappa) \rangle]$ using J = 0 and J' = 1, and a purely real mixing parameter $\chi(1)_0$ [13]. Remaining components of the dipole are enabled by J' = 1, M' = + 1, and they are $\langle t^1_\xi \rangle = \sqrt{2} \{\chi(1)_1'/\chi(1)_0\} \langle t^1_\zeta \rangle$ and $\langle t^1_\eta \rangle = \sqrt{2} \{\chi(1)_1''/\chi(1)_0\} \langle t^1_\zeta \rangle$, where a single prime and double prime denote the real and imaginary parts of the mixing parameter $\chi_1$, respectively. All foregoing results for $\langle t^1 \rangle$ are exact, and the form factor $[\langle j_0(\kappa) \rangle - (3/4) \langle j_2(\kappa) \rangle]$ is common to components of the Néel vector.

Next in line is a quadrupole and another valuable property, namely, multipoles of even rank are identically zero in a J-manifold. The $j_{eff}$ model of an iridate is a relevant example [3, 13]. The total angular momentum of an Ir ion is J = 5/2 and the corresponding quadrupole $\langle t^2 \rangle = 0$. However, a realistic model of $Sr_2IrO_4$, say, allows for a distortion of the environment along the c axis and J = 3/2 contaminates the J = 5/2 state and $\langle t^2 \rangle$ is nonzero. Specifically, the quadrupole depends on the electronic position operator $\mathbf{n}$. The equivalent operator $[(\mathbf{S} \times \mathbf{n}) \mathbf{n}]$ for $t^2$ shows it is a measure the correlation between the spin anapole $(\mathbf{S} \times \mathbf{n})$ and orbital degrees of freedom [3, 13]. A quadrupole has five components labelled by projections Q = 0, ±1, ±2. Ruthenium site symmetry does not restrict Q, for there is no spatial awareness other than a centre of inversion symmetry that forbids parity-odd Dirac multipoles [13]. Diagonal Q = 0 components of axial multipoles of rank K in neutron diffraction are identically zero for even K and $d^4$. Moreover, the rank K obeys the selection rule K = J' in $d^4$ [13]. The atomic configuration of $Ru^{4+}$ includes J = 0 and J' = 2. We find the exact results $\langle t^2_{\pm 1} \rangle = \{i\chi(2)_1/4\} \sqrt{(5/7)} \langle j_2(\kappa) \rangle$, $\langle t^2_{\pm 2} \rangle = \{\chi(2)_2/\chi(2)_1\} \langle t^2_{\pm 1} \rangle$, where $\chi(2)_1$ and $\chi(2)_2$ are the mixing parameter for M' = + 1 and M' = + 2, respectively, in the state J' = 2. In subsequent work, we restrict attention to dipoles and quadrupoles, setting aside octupoles and hexadecapoles allowed in $d^4$ amplitudes. The octupole form factor using J = 0 and J' = 3 is $[\langle j_2(\kappa) \rangle + (3/8) \langle j_4(\kappa) \rangle]$, and a hexadecapole is proportional to $\langle j_4(\kappa) \rangle$ [13].

The amplitude of magnetic neutron diffraction $\langle \mathbf{Q}_\perp \rangle = [\langle \mathbf{Q} \rangle - \mathbf{e} (\mathbf{e} \cdot \langle \mathbf{Q} \rangle)]$ yields an intensity $|\langle \mathbf{Q}_\perp \rangle|^2 = |\langle \mathbf{Q} \rangle|^2 - |(\mathbf{e} \cdot \langle \mathbf{Q} \rangle)|^2$, where the unit vector $\mathbf{e} = \boldsymbol{\kappa}/\kappa$ [9, 10]. Factors required to relate $|\langle \mathbf{Q}_\perp \rangle|^2$ to the measured intensity of a Bragg spot can be found in Ref. [14]. We use a shorthand $c = \cos(\beta) = \cos(124.662°) = -0.57$ and $s = \sin(\beta) = +0.82$, where β is the obtuse angle of the monoclinic cell in Fig. 1. Basis forbidden reflections include $\boldsymbol{\kappa} = (0, 0, L_o)_t$ with odd $L_o$. In this case $\langle Q_\eta \rangle \approx 0$ and,

$$\langle Q_{\perp\xi} \rangle \approx c \{[c \langle t^1_\xi \rangle - s \langle t^1_\zeta \rangle] + f\}, \langle Q_{\perp\zeta} \rangle \approx -s \{[c \langle t^1_\xi \rangle - s \langle t^1_\zeta \rangle] + f\}. \qquad (2)$$

Here, $f = (2/\sqrt{3}) [c \langle t^2_{+1} \rangle'' - s \langle t^2_{+2} \rangle'']$, with $\langle t^2_{+1} \rangle'' \propto (\eta\zeta)$ and $\langle t^2_{+2} \rangle'' \propto (\xi\eta)$ for quadrupole spatial awareness obtained from a spherical harmonic of rank 2. Amplitudes in Eq. (2) contain axial dipoles that form the Néel vector, and the same is true of all basis-forbidden amplitudes, i.e., the ferromagnetic component using $\langle t^1_\eta \rangle$ is not observed at basis-forbidden reflections with monoclinic Miller indices $k + l = 2m + 1$.

Berlijn et al. [15] did not find measurable intensities for reflection vectors $(0, 0, 1)_t$ and $(0, 0, 3)_t$ and conclude that dipole moments are parallel to the c axis of rutile ($P4_2'/mnm'$). Setting f = 0 in Eq. (2), and c $\langle t^1\xi\rangle$ = s $\langle t^1\zeta\rangle$ for null amplitudes places $\langle \mathbf{t}^1\rangle$ parallel to the crystal c axis. Inclusion of quadruples changes the result for the orientation of the axial dipole. Returning to Eq. (2), a null intensity implies {[c $\langle t^1\xi\rangle$ − s $\langle t^1\zeta\rangle$] + f} = 0, and in the corrected theory of diffraction $\langle \mathbf{t}^1\rangle$ subtends an angle ≈ − s (f/$\langle t^1\zeta\rangle$) to the c axis. Evidently, a significant factor in the magnitude of the deflection from the c axis is the ratio $\langle j_2(\kappa)\rangle/\langle j_0(\kappa)\rangle$ for which we find the value 0.193 at $(0, 0, 1)_t$. With regard to $(0, 0, 3)_t$, we see from Fig. 2 that $\langle j_0(\kappa)\rangle \approx 0$, to a good approximation, while $\langle j_2(\kappa)\rangle$ = 0.16. Null $(0, 0, 3)_t$ intensity infers that the projection of $\langle \mathbf{t}^1\rangle$ on the c axis is {(1 − s p)/√[1 + p (p − 2s)]} with p = (3f/$\langle j_2(\kappa)\rangle$ $\langle L\zeta\rangle$) and non-zero orbital angular momentum. An equally viable interpretation of a null $(0, 0, 3)_t$ intensity, and $\langle j_2(\kappa)\rangle$ different from zero, is $\chi(2)_r$ = 0 in the state J' = 2, and zero orbital angular momentum from $\chi(1)_r$ = 0.

Moving on, Berlijn et al. observed magnetic intensity at reflections $(1, 0, 0)_t$ and $(3, 0, 0)_t$ [15]. For $\boldsymbol{\kappa} = (0, k, 0)_m \equiv (− k, 0, 0)_t$ with k = 2m + 1 we obtain relatively simple amplitudes,

$$\langle Q_\perp\xi\rangle \approx \langle t^1\xi\rangle − (2/\sqrt{3}) \langle t^2_{+1}\rangle'', \quad \langle Q_\perp\eta\rangle \approx 0, \quad \langle Q_\perp\zeta\rangle \approx \langle t^1\zeta\rangle − (2/\sqrt{3}) \langle t^2_{+2}\rangle'', \quad (3)$$

since $(\mathbf{e} \cdot \langle \mathbf{Q}\rangle)$ = 0. Amplitudes Eq. (3) await a test. Individual components of $\langle \mathbf{Q}_\perp\rangle$ can be selected for observation in the spin-flip signal (SF) = {|$\langle \mathbf{Q}_\perp\rangle$|$^2$ − |$\mathbf{P} \cdot \langle \mathbf{Q}_\perp\rangle$|$^2$}, where neutron polarization $\mathbf{P}$ is assumed to be perfect. And, for future experiments using $\boldsymbol{\kappa} = (0, 0, l)_m \equiv (0, l, 0)_t$ with l = 2m + 1,

$$\langle Q\xi\rangle \approx \langle t^1\xi\rangle + (2/\sqrt{3}) \text{ s } [\text{s } \langle t^2_{+1}\rangle'' + \text{c } \langle t^2_{+2}\rangle''], \quad \langle Q\eta\rangle \approx 0,$$

$$\langle Q\zeta\rangle \approx \langle t^1\zeta\rangle + (2/\sqrt{3}) \text{ c } [\text{s } \langle t^2_{+1}\rangle'' + \text{c } \langle t^2_{+2}\rangle''], \quad (4)$$

with $(\mathbf{e} \cdot \langle \mathbf{Q}\rangle)$ = s $\langle t^1\zeta\rangle$ − c $\langle t^1\xi\rangle$. Quadrupoles in Eqs. (2) and (4) occur in different combinations.

*Chiral signature.* The magnetic symmetry considered for $RuO_2$ permits coupling with circular polarization in the primary beam of x-rays [16]. The coupling equates to a chiral signature for the symmetry and it demands a centrosymmetric crystal, magnetic order that does not break translation symmetry, and absence of anti-inversion ($\bar{1}'$) in the magnetic crystal class. Note that the latter demand rules out a linear magnetoelectric effect as a material property. A symmetry analysis of a parity-even absorption event shows that the corresponding chiral signature is caused by an interference between charge-like (time-even) and magnetic (time-odd) multipoles. At the level of accuracy to which we work, the signal is an interference between dipoles and charge-like quadrupoles, which create Templeton-Templeton scattering. In practice, the signature is captured in a difference between Bragg spot intensities measured with opposite handed primary x-rays.

Our results for resonant x-ray diffraction amplitudes exploit universal expressions [17]. In line with standard practice, primary photon polarizations labelled σ and π are perpendicular and parallel to the plane of scattering, respectively, and secondary polarization carry a prime [18-21]. Regarding diffraction amplitudes, (π'σ) and (σ'σ) apply to primary σ polarization rotated to the π channel and returned to the σ channel, respectively. In the diffraction setting crystals are rotated around the reflection vector by an angle ψ (an azimuthal angle scan). In contrast to neutron diffraction, the fixed x-ray energy severely limits the number of Bragg spots. Ruthenium $L_2$ and $L_3$ absorption edges occur at energies E ≈ 2.97 keV and E ≈ 2.84 keV, respectively, and a wavelength ≈ 12.4/E Å. Diffraction of x-rays by ruthenium multipoles with enhancement by an electric dipole - electric dipole (E1-E1) absorption event is described in terms of multipoles $\langle \mathbf{T}^K \rangle$ with rank K = 0, 1, 2. They have a time signature $(-1)^K$, and energy-integrated intensities satisfy sum rules at $L_2$ and $L_3$ edges [21, 22].

Bragg spots $(0, 1, 0)_t$ or $(0, 0, 1)_m$ correspond to a reflection vector $c^*_m = (2\pi/a)\,(0, 1, 0)$ parallel to the crystal b axis (cell length a ≈ 4.497 Å [15]). At the start of an azimuthal angle scan $b^*_m = (2\pi/a)\,(-1, 0, 0)$ is in the plane of scattering. E1-E1 amplitudes and the chiral signature are neatly expressed in terms of four purely real quantities,

$$A_1 = \sqrt{2}\,[c\,\langle T^1_\zeta \rangle + s\,\langle T^1_\xi \rangle],\ B_1 = \sqrt{2}\,[s\,\langle T^1_\zeta \rangle - c\,\langle T^1_\xi \rangle],$$

$$A_2 = 2\,[c\,\langle T^2_{+1} \rangle'' - s\,\langle T^2_{+2} \rangle''],\ B_2 = 2\,[s\,\langle T^2_{+1} \rangle'' + c\,\langle T^2_{+2} \rangle''].$$

Note that dipole and quadrupoles here and in neutron diffraction are arranged in different ways. The four amplitudes are (the crystal setting is different from [23]),

$$(\sigma'\sigma) = A_2 \sin(2\psi),\ (\pi'\pi) = i \sin(2\theta)\cos(\psi)\,A_1 + \sin^2(\theta)\sin(2\psi)\,A_2, \qquad (5)$$

$$(\pi'\sigma) = -i\cos(\theta)\sin(\psi)\,A_1 + i\sin(\theta)\,B_1 + \sin(\theta)\cos(2\psi)\,A_2 - \cos(\theta)\sin(\psi)\,B_2.$$

A change in sign of $A_1$ and $A_2$ relate (π'σ) and (σ'π). The Bragg angle for $(0, 0, 1)_m$ is sin(θ) ≈ 0.485 at the $L_3$ edge. A chiral signature ϒ requires all four amplitudes [Eq. (S2) in Supplemental Material [7]],

$$\Upsilon(c^*_m) = \cos(\theta)\cos(\psi)\,[\sin(2\theta)\sin(\psi)\,(A_1 B_2 - B_1 A_2)$$

$$- 2 A_1 A_2 \{\cos^2(\theta)\sin^2(\psi) + \sin^2(\theta)\}].\ (0, 0, 1)_m \qquad (6)$$

As already mentioned, interference between dipoles ($A_1$, $B_1$) and time-even quadrupoles ($A_2$, $B_2$) creates ϒ. By definition, the chiral signature expresses different phases between the four amplitudes, i.e., it vanishes if all four are purely real or purely imaginary.

The chiral signature for a reflection vector $b^*_m$ parallel to the crystal a axis, with Miller indices $(1, 0, 0)_t$ or $(0, −1, 0)_m$, is significantly different from Eq. (6) [23]. A key factor making the difference is that one $(0, −1, 0)_m$ scattering amplitude is zero. The result in question $(\sigma'\sigma) = 0$ for a reflection vector $b^*_m$ is a specific test of the proposed magnetic symmetry

*Conclusion.* In summary, we studied a magnetic symmetry for $RuO_2$ that accommodates a Néel vector that is not parallel to the rutile c axis, demanded by magnetic rutile symmetry ($P4_2'/mnm'$). In consequence, it fulfils a theory of anomalous thermal transport coefficients [6]. Magnetic symmetry $P2_1/c$ chosen for study permits a chiral signature in the diffraction of circularly polarized x-rays. Another acid test is the prediction of null Bragg intensity $(0, −1, 0)_m$ for unrotated photon polarization. Neutron diffraction amplitudes Eqs. (2), (3) and (4) await tests, as do the x-ray chiral signature Eq. (6) and diffraction amplitudes Eq. (5).

Regarding neutron diffraction, a zero magnetic quadrupole and zero orbital angular momentum are consistent with a viable interpretation of published data [15]. Specifically, null coupling constants between a state with total angular momentum $J = 0$ in the $Ru^{4+}$ ground state and $J' = 1, 2$, i.e., $\chi(J')_r = 0$, noting that the multipole rank = $J'$. A zero magnetic quadrupole is alien to altermagnetism, because the result implies a strong spin-orbit coupling - the likes of which is found in iridates, for example [3, 24]. The incomplete theory is probably justified in some cases, but it is definitely not a general concept, and in each case, one needs to decide to what extent the approach is applicable. With electronic spin degrees of freedom in altermagnetism unaware of the material environment they occupy it is a theory of magnetism fit for the ether. Our exact neutron form factors for the atomic configuration $d^4$ and exact diffraction amplitudes enable a peerless confirmation of $P2_1/c$ magnetic symmetry of $RuO_2$, or its emphatic rejection.

**References.**

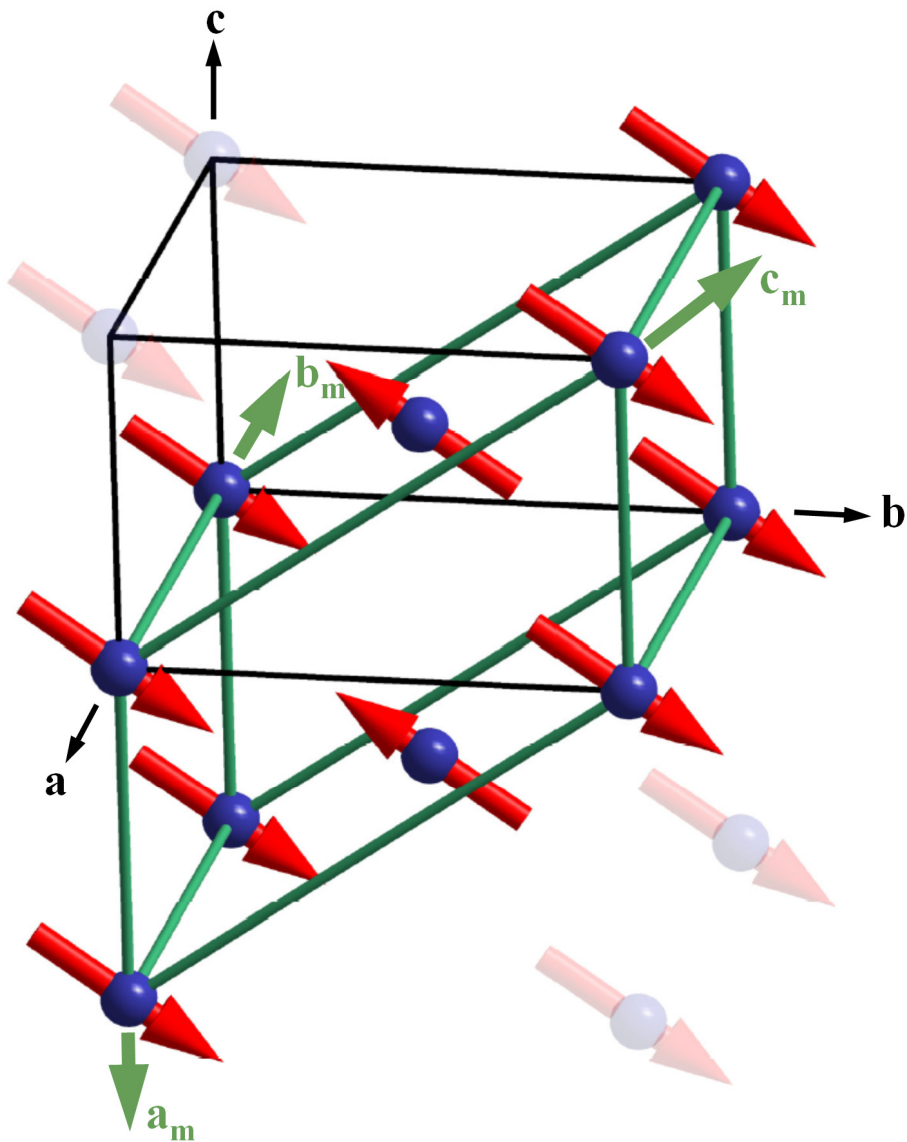

FIG. 1. Tetragonal (P4$_2$/mnm) and monoclinic (P2$_1$/c) unit cells. Vectors are defined in SI. Magnetic properties are referred to orthogonal vectors (ξ, η, ζ) that match (a$_m$*, b$_m$, c$_m$) where **a**$_m$* is a reciprocal lattice vector ∝ (**b**$_m$ × **c**$_m$).

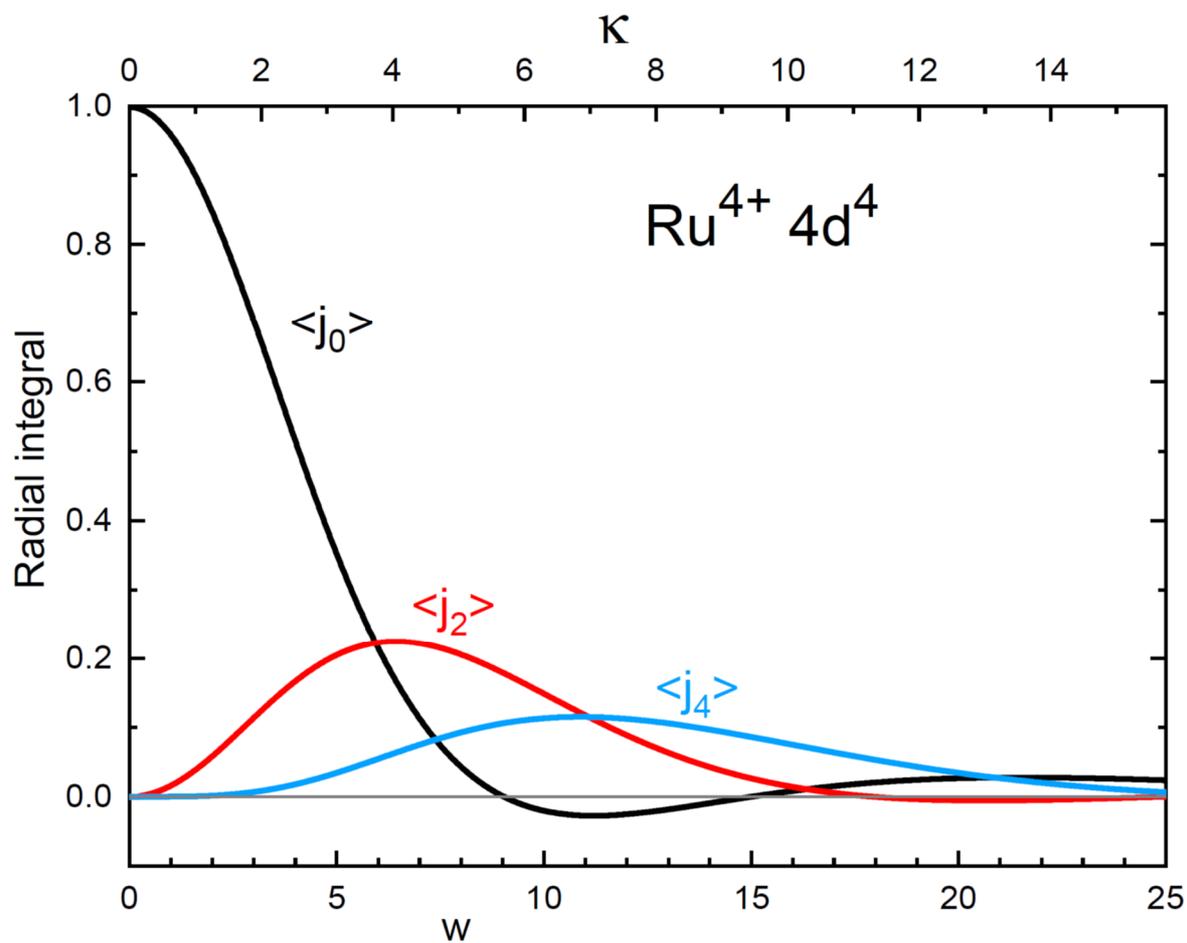

FIG. 2. Radial integrals $\langle j_0 \rangle$ (black), $\langle j_2 \rangle$ (red), and $\langle j_4 \rangle$ (blue) for $Ru^{4+}$ ($4d^4$) calculated using Cowan's code [11, 12]. The dimensionless parameter w and wavevector κ are related by the Bohr radius, namely, $w = 3a_o\kappa$.